**Bugs and features. A reply to Smaldino et al. (2017).**

Alberto Acerbi[1] & Edwin J. C. van Leeuwen[2]

[1] School of Innovation Science, Eindhoven University of Technology

[2] School of Psychology & Neuroscience, University of St Andrews

A conformity bias is, in cultural evolutionary terms, an individual-level tendency to disproportionately copy the majority. Collectively, at population-level, this individual behaviour results in a characteristic sigmoidal relationship between the frequency of the behaviour in the population and the probability to copy it ("the sigmoid"). In two recent papers (Acerbi et al. 2016, van Leeuwen et al. 2016), we, among other things, presented a series of simple models showing that, in some conditions, population-level outcomes produced without a conformist bias could also generate a sigmoid ("generate a sigmoid" here means that, when fitting the data both with a linear and a sigmoid model, the latter provides a better fit). We did not claim that the sigmoid is *never* evidence for an individual-level conformist bias, but that empiricists might want to pay attention to the possibility that mechanisms other than conformity can explain its presence. More generally, we invited to be cautious when inferring individual-level behaviours from population-level patterns.

In particular, we identified three scenarios in which the sigmoid could be generated in the absence of conformity bias:

(1) When there is a preference for one cultural trait over the other (the condition "*Variant Preference*" in Acerbi et al. 2016).

(2) When the (random) choice of demonstrators for individuals to copy is limited to a subset of the population (the condition "*Demonstrators subgroup*" in Acerbi et al. 2016).

(3) When copying probability is plotted against the cumulative frequency of behaviours in the population, instead of against the frequency of individuals showing the behaviour at each time step (van Leeuwen et al. 2016).

In Acerbi et al. (2016) and van Leeuwen et al. (2016) we provided more details about the implementations, and about real-life situations in which we expect these scenarios to be relevant (note that in Acerbi et al. 2016 we also found six scenarios that do *not* produce a sigmoid as an outcome, so that one can exclude real-life instances of those cases as possible confoundings for the conformity interpretation).

Smaldino, Aplin & Farine (2017) provide a thorough criticism of our models, suggesting that "problematic assumptions" and "implementation errors" make our "overly strong claims" unlikely to be correct. We take in all seriousness the criticism of Smaldino et al. – which, we believe, is formally correct – but we also think that their analysis, while possibly clarifying aspects of the models overlooked in our papers, mostly corroborates the importance of our findings. Below we address separately their criticisms of our three scenarios.

Regarding the scenario "*Variant Preference*", Smaldino et al. show that the sigmoidal curve is produced (in part) because the outputs of the simulations are averaged across all runs, and the simulations are initialised by randomly populating the individuals with one of the two variants. This is correct, but we do not believe these assumptions are problematic. The first assumption – averaging the runs of the simulations – was a feature of our model, not a bug. We indeed intended to show that experiments in which results of different runs are pooled together (as in the key empirical study considered by Smaldino et al., namely, Aplin et al. 2015) might produce the sigmoidal outcome in the absence of conformity, when a stable within-run preference for one of the two variants is present (see Acerbi et al. 2016 for a discussion of some realistic possibilities where this may happen).

The second assumption – initialising randomly each individual in the population with one of the two variants – was indeed an arbitrary choice, but it is a standard practice in modelling. Moreover, there are several empirical studies in which conformity was tested after many, if not all, individuals had obtained one or the other cultural variant (e.g. Whiten, Horner & de Waal 2005; Dindo, Whiten & de Waal 2009; Perry 2009; Wrangham et al. 2017) so that our choice and subsequent results are relevant to empiricists. In addition to this, where we suspected that this way of initialising the population was the main reason for the results obtained, we analysed an alternative scenario mimicking studies in which the spread of cultural variants from a small part of the population was analysed for conformist patterns (e.g. Aplin et al. 2015). This model, absent from Smaldino et al.'s analysis and discussion, produced similar results (van Leeuwen et al., 2016, and see below, in the discussion of our third scenario). What would be the effects of this alternative set-up in the "Variant Preference" condition is an open question.

Smaldino et al.'s "disagreement" with respect to our second scenario ("*Demonstrators subgroup*") is in fact completely in line with our own interpretation of the results, and may count as a successful replication of our model. The fact that, to produce a sigmoid in absence of conformity at individual-level, the total group size and the total number of demonstrator should be relatively small is studied in detail in our own paper (the caption of Figure 4 in Acerbi et al. 2016 reads: "Sigmoid fit is more evident in small populations in the condition Demonstrators subgroup"). Smaldino et al. thus corroborate our important finding that conformity cannot be identified based on the population-level sigmoid when the alternative mechanism of "copy-any-kind-of-subset of the (relatively small) population" (e.g., the dominants, old-aged, or experts) cannot be ruled out.

Finally, Smaldino et al.'s criticism of our third scenario is not based on a disagreement on the model's implementation. They agree that, even with unbiased transmission, when counting the cumulative frequency of behaviours instead of the number of individuals showing the behaviour at each time step, a sigmoidal outcome will be produced. However, Smaldino et al. note correctly that this is also due to our decision to sample the entire history of previous events of the agents, which they evaluate as an approach "highly unlikely to be used by empiricists".

Regarding the first aspect – counting behaviours instead of individuals – we discussed elsewhere why we think that this a problematic approach (see e.g. Leeuwen et al. 2015), and our reason to model it was indeed to show other problems that could arise in experiments adopting such methodology (e.g. Claidière, Bowler, & Whiten 2012, Claidière et al. 2014, Aplin et al. 2015).

Regarding the second aspect – sampling from the entire history of events – we acknowledge that applying a limited time-window to the calculation of the observed distribution of traits may be sensible, especially given Smaldino et al.'s result that restricted lengths of counted history cause the sigmoid to dissolve into a linear relationship. At the same time, we note that, to our knowledge, there are no studies purporting evidence of conformity other than Aplin et al. (2015) that apply any time-window for the calculation of observed trait distributions. In other words, our model applies to other conformity studies and Smaldino et al.'s results should be taken as a validation of our cautionary statements in this regard. Note also that the time window of ~2,000 events that Smaldino et al. identify as the threshold for producing the sigmoid in absence of conformity is not that large, as events represent individual interactions. For an experimenter sampling a population of 200 individuals, for example, 10 copying events for each subject would be enough. The correct comparison, thus, is not between the 6 "solves" of Aplin et al. (2015) and the 2,000 events identified by the analysis, as implied in Smaldino et al., but between the 6 "solves" *times the total population of the experiment* and the 2,000 events.

In addition, Smaldino et al. correctly demonstrate that our main result is also due to the initialisation of the population with random behaviours, while in experiments generally only one individual shows the tracked variant at the beginning of each run. As we mentioned above, we also recognised that this is what mainly explained the sigmoid in this model. In fact, in van Leeuwen et al. (2016), we discuss this very point, and we present an additional model, in which simulations start with a population of naïve individuals, and behaviours are introduced by individual innovations. The only twist is that innovation rate decreases over time. The outcomes of this model are analogous to the random-initialisation case (see van Leeuwen et al. 2016), and validate, in our view, the approach we took.

In sum, the fact that our models generate "artefactual" sigmoidal curves (i.e. in the absence of conformity) is indeed what we wanted to demonstrate. Our "faulty" assumptions were intended as showing that, *if similar assumptions are made in observations and experiments*, they would produce sigmoidal curves in the absence of an individual-level conformist bias. Smaldino et al. think that this is unlikely, while we think it might be worth to keep this in mind.

We invite the interested readers to scrutinize our – and Smaldino et al. 's – full papers to make up their own mind about the arguments that we here just sketched. We hope this exchange can show the utility of threaded scientific discussion, as well as of open science (all our codes were made available at the time of publication – Acerbi, 2016). Above all, we believe that this exchange highlights the importance of modelling practices – as opposed to purely verbal theorising – in social and biological sciences. Only with models it is possible to check assumptions quantitatively, evaluate their realism and scope, and, if necessary, change and improve them.